\documentstyle[preprint,aps]{revtex}
\begin{document}
\draft
\title{Learning and Generalization Theories of Large Committee--Machines}
\author{R\'emi Monasson\cite{rm} and Riccardo Zecchina\cite{rz}}
\address{\cite{rm} Laboratoire de Physique Th\'eorique de l'ENS,
24 rue Lhomond, 75231 Paris cedex 05, France\\
\cite{rz} INFN and Dip. di Fisica, Politecnico di Torino,
C.so Duca degli Abruzzi 24, I-10129 Torino, Italy}
\date{December 1995}
\maketitle

\begin{abstract}
The study of the distribution of volumes associated to the internal
representations of learning examples allows us to derive
the critical learning capacity ($\alpha_c=\frac{16}{\pi} \sqrt{\ln K}$) 
of large committee machines, to verify
the stability of the solution 
in the limit of a large number $K$ of hidden units and to find
a Bayesian generalization cross--over at $\alpha=K$.
\end{abstract}


\section{Introduction}

Following the approach presented in refs.\cite{nosotros,rete}, we derive
the learning behaviour of non overlapping committee machines with a
large number $K$ of hidden units. Scope of the paper is to clarify some of
the analytical aspects of a method which is based on the internal degrees of
freedom of MultiLayer Networks (MLN) and which requires a double analytic
continuation. Such an approach, beside allowing for
the derivation of new results both for the learning and the generalization
behaviour of MLN, makes a rigorous bridge between different fields in the
theory of neural computation, such as Information Theory, VC--dimension and
Bayesian rule extraction, and statistical mechanics 
\cite{nosotros,mit,vc,opper,watkin}. Moreover, it sheds new
light on the role of internal representations of the learning examples by
relating it to the distribution of domains of solutions and pure states in
the weight space of the network. 

The method consists in a generalization of the well known Gardner 
approach\cite{gard}.
While the latter studies the typical volume of couplings associated to the
overall input--output map implemented by the network,
here we consider the decomposition of such volume in
a macroscopic number of single volumes associated to all 
possible internal representations compatible with the learned examples.
In addition to the interaction weights, we take as dynamical variables 
also the internal 
state variables of the MLN  characterizing internal representations.
For the storage problem, we are therefore interested in counting the typical 
number 
$\exp({\cal N}_D)$ of volumes giving the dominant contribution to
Gardner's volume and to compare it with the total number $\exp({\cal N}_R)$
of non--empty volumes. At the learning transition, i.e. when the Gardner's
total volume shrinks to zero and no more patterns can be learned without
errors, we expect both entropies ${\cal N}_D$ and ${\cal N}_R$ to vanish. 
The vanishing condition on ${\cal N}_D$  and ${\cal N}_R$ gives thus
an alternative indication on the storage performance of the studied network.

The generalization properties of the network, i.e. the rule inference
capability from a given set of deterministic input--output examples,
also depend on the geometrical structure of the weight space.
As we shall discuss in the sequel, the internal representation approach
can be straightforwardly extended to the study of the geralization error
of MLN in the Bayesian framework \cite{watkin,opper}, allowing 
for a geometrical
interpretation of the generalization transition together with a clarification
of the role of the VC-dimension\cite{vc}.

The method discussed here for the committee machine can be
straightfordwardly extended 
to other non overlapping MLN with arbitrary
decoder functions. For instance, one may show that the stability analysis of
the RS solution $\alpha _c=\ln K/\ln 2$ for parity machine is an exact
result in the limit $K>>1$. Such a result coincides with the one derived in 
\cite{bhs} following the standard Gardner \cite{gard} approach with one step
of Replica Symmetry Breaking. Moreover, one may also reproduce the known 
results\cite{opper} on the parity machine generalization 
transition\cite{nosotros} by means of a detailed geometrical intepretation.

The paper is organized as follows.
In Sec.{\bf II}  we outline the basic points of our approach, both
for learning and for the Bayesian generalization problems. 
In Sec.{\bf III} and Sec.{\bf IV} we study the entropies
${\cal N}_R$ and ${\cal N}_D$ in the $K>>1$ limit  
and compute the closed expression for
the critical capacity. The detailed analysis of the stability of 
the solution is given in Sec. {\bf V}.
Finally, in Sec.{\bf VI}, we derive the entropy ${\cal M}_D$ (for $K>>1$)
of the internal representation contributing to the Bayesian entropy. This
allows us to analyze the generalization transistion of the committee
machine and to explain why the VC--dimension is not relevant for its 
typical generalization properties. 
 
\section{The Internal Representation Volumes Approach}

We consider tree-like commitee machines composed of $K$ non-overlapping
perceptrons with real-valued weights $J_{\ell i}$ and connected to $K$ sets
of independent inputs $\xi _{\ell i}$ ($\ell =1,...,K$, $i=1,...,N/K$).
Committee machines are characterized by an output $\sigma$ which is a binary
function $f(\{\tau_\ell\})=\hbox{sign}(\sum_\ell \tau_\ell)$ of the cells $%
\tau_\ell = \hbox{sign} ( \sum _i J_{\ell i} \xi _{\ell i})$ in the hidden
layer. We refer to the set $\{ \tau _\ell \}$ as the {\em internal
representation} of the input pattern $\{ \xi _{\ell i} \}$. Given a
macroscopic set of $P=\alpha N$ binary unbiased patterns patterns (the
training set), the learning problem consists in finding a suitable set of
internal representations ${\cal T } = \{ \tau _\ell ^{\mu} \}$ with a
corresponding non zero volume 
\begin{equation}
V_{{\cal T }} = \int \prod_{\ell ,i} dJ_{\ell i} \prod_{\mu} \theta \left(
\sigma ^{\mu} f(\{\tau_\ell ^\mu\}) \right) \prod _{\mu ,\ell} \theta \left(
\tau _\ell ^{\mu} \sum _i J_{\ell i} \xi _{\ell i} ^{\mu} \right) \; \; ,
\; \; \int \prod_{\ell ,i}dJ_{\ell i} = 1 \; \; \;,
\label{volume}
\end{equation}
where $\theta(\dots)$ is the Heaviside function.

The total volume of the weight space available for learning, i.e. Gardner's
total volume, is given by $V_G=\sum_{{\cal T}}V_{{\cal T}}$. We are
interested in discussing the limit $\overline{\ln V_G}\to -\infty$, 
which defines
the maximal possible size of the training set or the critical capacity of
the model. The bar denotes the average over the patterns and their
corresponding outputs \cite{gard} which, as usual, are drawn according to
the binary unbiased distribution law.

As discussed in \cite{nosotros}, the partition of $V_G$ into connected
components may be naturally obtained using the volumes $V_{{\cal T}}$
associated to the internal representations. This allows to give a
geometrical interpretation of the learning and Bayesian generalization
process in terms of the characteristics of the volumes dominating the
overall distribution\cite{nosotros}.

Following the standard statistical mechanics approach, we first compute 
\begin{equation}
g(r)\equiv - \frac{1}{Nr}
\overline{\ln \left(\sum_{{\cal T}} V_{{\cal T}}
^{ {\displaystyle r}} \right) } \label{gdef}
\end{equation}
and next derive the entropy ${\cal N}(w)$ of the volumes $V_{{\cal T}}$
whose sizes are equal to $w={\frac 1N}\ln V_{{\cal T}}$. This can be done
using the Legendre relations $w_r={\frac{\partial (rg(r))}{\partial r}}$ and 
${\cal N}(w_r)=-{\frac{\partial g(r)}{\partial (1/r)}}$ . Diversely from the
standard replica calculations, here we deal with two analytic continuations:
we have $r$ blocks of $n$ replicas and, once the we average over the
quenched patterns for $r$ and $n$ integer has been done, we perform an
analytic continuation to real values of $r$ and $n$. Labeling blocks and
replicas by $\rho ,\lambda $ and $a,b$ respectively, the spin glass order
parameters read 
\begin{equation}
q_\ell ^{a\rho ,b\lambda }=\frac KN\sum_iJ_{i\ell }^{a\rho }J_{i\ell
}^{b\lambda }\,\,\,\,\,\,.
\end{equation}
They represent the typical overlaps between weight vectors incoming onto the
same hidden unit $\ell $ ($\ell =1,\dots ,K$) and belinging to blocks $\rho
,\lambda $ and replicas $a,b$. Associated to the the $q_\ell
^{a\rho ,b\lambda }$ there are also the conjugate Lagrange multipliers ${%
\hat q_\ell }^{a\rho ,b\lambda }$. Since hidden units are equivalent, we
assume that at the saddle point $q_\ell ^{a\rho ,b\lambda }=q^{a\rho
,b\lambda }$ and ${\hat q_\ell }^{a\rho ,b\lambda }={\hat q}^{a\rho
,b\lambda }$ independently of $\ell $. Then, within the replica symmetric
(RS) Ansatz \cite{mpv}, we find 
\begin{eqnarray}
g(r)=\mathop{\rm Extr}_{q,q_{*}}\Bigg\{ &&{\frac{1-r}{2r}}\ln (1-q_{*})-{%
\frac 1{2r}}\ln (1-q_{*}+r(q_{*}-q))-{\frac q{2(1-q_{*}+r(q_{*}-q))}} 
\nonumber \\
&&-{\frac \alpha r}\int \prod_{\ell=1}^K Dx_\ell \ \ln {\cal H}(\{x_\ell \})%
\Bigg\} \;\;\;,  \label{g_r}
\end{eqnarray}
where 
\begin{equation}
{\cal H}(\{x_\ell \})=\mathop{\rm Tr}_{\{\tau _\ell \}}\prod_{\ell=1}^K \int
Dy_\ell H\left[\frac{y_\ell \sqrt{q_{*}-q}+\tau _\ell x_\ell \sqrt{q}}
{\sqrt{1-q_{*}}} \right]^r\;\;\;.  \label{Tr}
\end{equation}
Here, $q_{*}(r)=q^{a\rho ,a\lambda }$ and $q(r)=q^{a\rho ,b\lambda }$ are
the typical overlaps between two weight vectors corresponding to the same ($%
a $,$\rho \ne \lambda $) and to different ($a\ne b$) internal
representations ${\cal T}$ respectively \cite{gard,rete}. The Gaussian
measure is denoted by $Dx=\frac 1{\sqrt{2\pi }}e^{-x^2/2}$ whereas the
function $H$ is defined as $H(y)=\int_y^\infty Dx$. Since with no loss of
genereality the outputs $\sigma ^\mu $ can be set equal to $1$, in eqn.(\ref
{g_r}) the sum $\mathop{\rm Tr}_{\{\tau _\ell \}}$ runs over the internal
representations $\{\tau _\ell \}$ giving a positive output $f(\{\tau _\ell
\})=+1$ only.

As discussed in \cite{nosotros}, when $N\to \infty $, $\frac 1N\overline{%
\ln (V_G)}=-g(r=1)$ is dominated by volumes of size $w_{r=1}$ whose
corresponding entropy (i.e. the logarithm of their number divided by $N$) is 
${\cal N}_D={\cal N}(w_{r=1})$. At the same time the most numerous volumes
are those of smaller size $w_{r=0}$, since in the limit $r\to 0$ all the $%
{\cal T}$ are counted irrespectively of their relative volumes. Their
corresponding entropy ${\cal N}_R={\cal N}(w_{r=0})$ is the (normalized)
logarithm of the total number of implementable internal representations. The
quantities ${\cal N}_D$ and ${\cal N}_R$ are easily obtained from the RS
free--energy eqn.(\ref{g_r}) using Legendre identities. In particular, $%
q(r=1)$ is the usual saddle point overlap of the Gardner volume $g(1)$ \cite
{gard,bhs}. The vanishing condition for the entropies is related to the
zero volume condition for $V_G$ and thus to the storage capacity of the
models.

In the above discussion, we have focused on the storage problem.
However, the generalization properties also depend on the internal structure
of the coupling space. Let us for instance consider the case of a learnable
rule, defined by a teacher network. When a student with the same architecture
is given more and more examples of the rule to infer, its version space
\cite{watkin} shrinks. In the perceptron case, the version space
is simply connected and the typical generalization error done by the
student on a new example goes to zero as its overlap with the teacher
increases. The situation is much more involved in multilayer neural networks
since the presence of separated components of the version space makes the
alignment of the student along the teacher direction more difficult.
The approach we have exposed above for the learning problem may be 
extended to acquire a better understanding of the generalization process in
multilayer networks.

We shall restrict to the Bayesian framework where all teacher are sorted
according to their a priori probabilities. The generalization properties
are derived through the Bayesian entropy
\begin{equation}
S_G = - \frac 1N \sum _{\{\sigma ^\mu\}} \overline{V_G \ln V_G} \; \; \;,
\end{equation}
where the sum runs over all $2^P$ sets of possible outputs.
If we know intend to look at the distibution of the sizes fo the internal
representation volumes $V_{{\cal T}}$, we have to
consider the generating free--energy\cite{corr}
\begin{equation}
s(r)\equiv - \frac{1}{Nr} \sum _{\{\sigma ^\mu\}}
\overline{V_G \ln \left(\sum_{{\cal T}} V_{{\cal T}}
^{ {\displaystyle r}} \right) } \; \; \;. 
\label{sdef}
\end{equation}
To compute $s(r)$ with the replica method, we have to introduce $1+nr$
replicas and send $n\to 0$ at the end of the computation. The order
parameters entering the computation are the overlaps $p^{a\rho}$ between the
teacher and the $nr$ students and the overlaps $q^{a\rho,b\lambda}$
between two differents students. Within the RS Ansatz, we assume that
$p^{a\rho}=p$ and $q^{a\rho,b\lambda}=q$ if $a\ne b$, $q^*$ otherwise.
The result we obtained is
\begin{eqnarray}
s(r)=\mathop{\rm Extr}_{p,q,q_{*}}\Bigg\{ &&{
\frac{1-r}{2r}}\ln (1-q_{*})-{%
\frac 1{2r}}\ln (1-q_{*}+r(q_{*}-q))-{\frac {q-p^2}{2(1-q_{*}+r(q_{*}-q))}} 
\nonumber \\
&&-\frac{2\alpha}{r}\int \prod_{\ell=1}^K Dx_\ell 
\left[ \mathop{\rm Tr}_{\{\tau _\ell \}} \prod _\ell H \left(
\frac{\tau_\ell x_\ell p}{\sqrt{q_0-p^2}}\right) \right]
\ln {\cal H}(\{x_\ell \})%
\Bigg\} \;\;\;,  \label{s_r}
\end{eqnarray}
In the following , we shall focus on the logarithm (divided by $N$) of the 
number of internal representations contributing to $S_G=-s(1)$, that is
\begin{equation}
{\cal M}_D = \frac{\partial s}{\partial r} (r=1) \; \; \;. 
\end{equation}
It can be easily verified that $p=q$ is always a saddle--point when $r=1$.
In the Bayesian framework, the typical overlap between two student is
equal to the scalar product between the teacher and any student.

\section{Analysis of ${\cal N}_R$ in the $K>>1$ Limit}

We first focus on the $r\to 0$ case to compute the typical logarithm 
${\cal N}_R$ of the total number of internal representations.
One can check on the saddle--point equations for $q,q^*$ that the
correct scalings of the order parameters are $q=O(1)$ and $q^*=1-O(r)$.
In the following, we shall call $\mu = \lim _{r\to 0} [r/(1-q^*)]$.
Upon keeping the leading terms in $K$, the trace over ${\cal T}$ in
equation (\ref{Tr}) becomes \cite{bhs}
\begin{equation}
H\left(\frac{Q_1}{\sqrt{1-Q_2}}\right)\ \prod_{\ell =1}^K \ A(x_\ell)
\end{equation}
where 
\begin{equation}
Q_1=\frac 1{\sqrt{K}}\sum_{\ell=1}^K \frac{B(x_\ell) +B(-x_\ell)}{A(x_\ell) }\;\;\;,
\end{equation}
and 
\begin{equation}
Q_2=\frac 1K\sum_{\ell=1}^K \left(\frac{B(x_\ell)+B(-x_\ell) }{A(x_\ell) }
\right)^2 \;\;\; .
\end{equation}
In the above expressions we have adopted the definitions 
\begin{equation}
A(x) \equiv 1+\frac{\exp \left(-x ^2\frac{\mu q}{2(1+\mu (1-q))}
\right)} {\sqrt{1+\mu (1-q)}}
\end{equation}
and 
\begin{equation}
B(x)\equiv H\left(x \sqrt{ \frac{q}{1-q}}\right) +
\frac{\exp \left(-x^2\frac{\mu q}{2(1+\mu (1-q))} \right)} 
{\sqrt{1+\mu (1-q)}} H\left(\frac{-x \sqrt{q/(1-q)}}{\sqrt{1+\mu (1-q)}}\right)
\; \; \; .
\end{equation}
In the $K\to \infty$ limit, $Q_1$ becomes a gaussian variable with zero mean
and variance $Q_2=\int Dx (B(x)+B(-x))^2/A(x)^2$. The free--energy for
$r\to 0$ then reads $-G(q,\mu)/r+O(\ln r)$ where
\begin{eqnarray}
G(q,\mu)&=& \frac{1}{2} \ln(1+\mu(1-q))+ \frac{1}{2} \frac{\mu q}
{1+\mu (1-q)}+
\alpha \int Dx \ln H\left(\frac{x \sqrt{Q_2}}{\sqrt{1-Q_2}} \right)+
\nonumber \\ & & \alpha K \int Dx
\ln\left( 1+\frac{\exp \left(-x^2 \frac{\mu q}{2 (1+\mu(1-q))}
\right)} {\sqrt{1+\mu(1-q)}}\right)
+O\left(\frac{1}{\sqrt{K}}\right)
\end{eqnarray}
and the typical logarithm ${\cal N}_R$ 
of the total number of internal representations
is simply the maximum of $G(q,\mu)$ over $q$ and $\mu$.
Taking the scaling relation $\mu= m K^2$ (which can be inferred from the
equation $\frac{\partial G}{\partial \mu}=0$), and $q=O(1)$, one finds 
\begin{equation}
Q_2=\frac{2}{\pi} \arcsin(q) \; \; \;.
\end{equation}
Finally, defining $q=1-\epsilon$ and taking the saddle point equation with
respect to $m$ (which implies $m=\alpha^2$) and $\epsilon$, one finds the
following result
\begin{equation}
{\cal N}_R=\ln (K)-\frac{\pi ^2\alpha^2}{256}+O(\ln(\alpha)) \; \; \;,
\end{equation}
which vanishes at 
\begin{equation}
\alpha_R=\frac{16}{\pi} \sqrt{\ln(K) }\; \; \;.
\end{equation}

\section{Analysis of ${\cal N}_D$ in the $K>>1$ Limit}

We shall now concentrate on the $r\to 1$ case, which corresponds to
the internal representations giving the dominant contribution to the
Gardner volume $V_G$. The typical logarithm of such internal representations
is ${\cal N}_D$. 
Before taking the limit $K\to \infty$, the Legendre transform of
expression (\ref{g_r}) gives 
\begin{eqnarray}
{\cal N}_D &=& \frac 1N \overline{\ln V_G} +\frac{2 q q^*-q^*-q^2}
{2(1-q)^2}-\frac 12 \ln (1-q^*) - \alpha K \times \nonumber \\
&& \int \prod_\ell Dx_\ell \frac{\mathop{\rm Tr}_{\{\tau _\ell \}}
\prod_{l=2}^K H\left( \tau_\ell x_\ell \sqrt{\frac{q}{1-q}} \right)
\int Dy H\left(\frac{y\sqrt{q^*-q}+x_1 \tau_1 \sqrt{q}}{\sqrt{1-q^*}}
\right) \ln H\left(\frac{y\sqrt{q^*-q}+x_1 \tau_1 \sqrt{q}}{\sqrt{1-q^*}}
\right) } {\mathop{\rm Tr}_{\{\tau _\ell \}} \prod_{l=1}^K H\left( 
\tau_\ell x_\ell \sqrt{\frac{q}{1-q}} \right) }
\end{eqnarray}
where $\overline{\ln V_G}$ is the replica symmetric expression of
the Gardner volume.
When $K$ is large, the trace over all allowed internal representations
may be evaluated as in the previous section\cite{bhs}. We find the
following scalings for the order parameters
\begin{eqnarray}
q &\simeq& 1- \frac{128}{\pi^2 \alpha^2} \nonumber \\
Q_2 &\simeq& 1- \frac{32}{\pi^2 \alpha} \nonumber \\
q^*  &\simeq& 1- \frac{\Gamma^2}{2 \pi^2 K^2 \alpha^2}
\end{eqnarray}
where 
\begin{equation}
\frac{1}{\Gamma} = - \sqrt \pi\ \int _{-\infty}^{\infty} \ du
H(u) \ln H(u) \label{Gam}
\end{equation}
for large $K$ and $\alpha$. Therefore, the asymptotic expression of the
entropy of contributing internal representations is
\begin{equation}
{\cal N}_D=\ln (K)-\frac{\pi ^2\alpha^2}{256}+O(\ln(\alpha)) \; \; \;,
\end{equation}
which vanishes at 
\begin{equation}
\alpha_D=\frac{16}{\pi} \sqrt{\ln(K) }\; \; \;.
\end{equation}
For large $K$, $\alpha_D$ coincide with $\alpha_R$. Reasonnably, we
expect $\alpha_c$ to be equal to these critical numbers and to scale
as $\frac{16}{\pi} \sqrt{\ln(K) }$ too.

\section{Stability Analysis}

In order to show that our RS calculation of ${\cal N}_R$ is 
asymptotically correct when the number of hidden units $K$ is large, we have
checked its local stability with respect to fluctuations of the order
parameter matrices.  Although it would require a complete analysis of
the eigenvalues of the Hessian matrix, we have focused only on the
replicons 011 and 122 in the notations of \cite{kondor}, which are
usually the most ``dangerous'' modes \cite{mpv}. For a free--energy
functional depending only on one order parameter matrix $q^{a\rho
,b\lambda}$ , the corresponding eigenvalues are
\begin{eqnarray}
\Lambda _{011} &=& \frac{\partial ^2 {\cal F}}{\partial q^{a\rho,b
\lambda} \partial q^{a\rho,b\lambda}}- 2r\frac{\partial ^2 {\cal
F}}{\partial q^{a\rho,b\lambda} \partial q^{a\rho,c\mu}}+
2(r-1)\frac{\partial ^2 {\cal F}}{\partial q^{a\rho,b\lambda} 
\partial q^{a\rho,b\mu}}+ \nonumber \\ & &(r-1)^2 \frac{\partial ^2 {\cal
F}}{\partial q^{a\rho,b\lambda} \partial q^{a\mu,b\nu}} - 2r(r-1)
\frac{\partial ^2 {\cal F}}{\partial q^{a\rho,b\lambda} \partial
q^{a\mu,c\nu}} + r^2\frac{\partial ^2 {\cal F}}{\partial
q^{a\rho,b\lambda} \partial q^{c\mu,d\nu}}
\end{eqnarray}
and
\begin{equation}
\Lambda_{122}=\frac{\partial ^2 {\cal F}}{\partial q^{a\rho,a\lambda}
\partial q^{a\rho,a\lambda}}- 2\frac{\partial ^2 {\cal F}}{\partial
q^{a\rho,a\lambda} \partial q^{a\rho,a\mu}}+ \frac{\partial ^2 {\cal
F}}{\partial q^{a\rho,a\lambda} \partial q^{a\mu,b\nu}}
\end{equation}
are given by formula (41) in ref.\cite{kondor}.  In our case, however,
the free-energy depends upon the $2K$ matrices $\{ {\cal
Q}_l,{\hat{\cal Q}}_l \}$. According to \cite{gard,bhs}, the stability 
condition for each mode reads
\begin{equation}
\Delta (\alpha ,K) = \hat \Lambda \ (\ \Lambda + (K-1) \overline{
\Lambda}\ ) - {1 \over K^2} < 0
\end{equation}
where $\hat \Lambda,\Lambda,\overline{\Lambda}$ are the eigenvalues
computed for the fluctuations with respect to ${\hat{\cal Q}}_\ell
{\hat{\cal Q}}_\ell $, ${\cal Q}_\ell {\cal Q}_\ell $ and ${\cal
Q}_\ell {\cal Q}_m$ ($\ell \ne m$) respectively.  Since we are
interested in the stability of the saddle--point giving ${\cal N}_R$,
we focus on the limit $r\to 0$. In this case, the correct scalings of
the order parameters are $q^*=1- r/\mu +O(r^2)$, $q=O(1)$. 
For the $(011)$ mode, we find 
\begin{eqnarray}
\hat \Lambda _{011} &=& \frac{(1-q)^2}{K} r^2 \nonumber \\
\Lambda _{011} &=& \frac{\alpha \mu ^2}{r^2} \int \prod_{\ell=1}^K Dx_\ell
\left[\frac{N_1}{D}-\mu \left( \frac{N_2}{D} \right) ^2\right] ^2 \nonumber \\
\overline{\Lambda}_{011} &=& \frac{\alpha \mu ^4}{r^2} \int \prod_{\ell=1}^K 
Dx_\ell \left[\frac{N_3}{D} -\frac{N_4}{D^2}\right]^2
\label{011}
\end{eqnarray}
at leading order when $r\ll 1$. 
The quantities defined in (\ref{011}) are
\begin{equation}
D=\mathop{\rm Tr}_{\{\tau _\ell \}} \prod _{\ell=1}^K B (\tau_\ell x_\ell)
\end{equation}
\begin{eqnarray}
N_1 &=& \mathop{\rm Tr}_{\{\tau _\ell \}} \left[ \prod _{\ell=2}^K 
B(\tau_\ell x_\ell)\right]
\frac{\exp \left(-x_1^2\frac{\mu q}{2 X )}
\right)} {X^{3/2}} \left[ \left(1-\frac{\mu q x_1^2}{X}
\right) \right. \times \nonumber \\
&& \left. H\left(\frac{-\tau_1 x_1 \sqrt{q/(1-q)}}{\sqrt{X}}\right)
- \frac{\mu \sqrt{q(1-q)}}{\sqrt{2\pi}\sqrt{X}} 
\exp \left(-\frac{x_1^2 \mu q}{2 X)(1-q)  }\right)\right] \nonumber \\
N_2 &=& \mathop{\rm Tr}_{\{\tau _\ell \}} \left[ \prod _{\ell=2}^K 
B(\tau_\ell x_\ell)\right] \frac{\sqrt{1-q}}{X} 
 Y_1
\nonumber \\
N_3 &=& \mathop{\rm Tr}_{\{\tau _\ell \}} \left[ \prod _{\ell=3}^K 
B(\tau_\ell x_\ell)\right] \frac{1-q}{X^2}
Y_1 Y_2 
\nonumber \\
N_4 &=& \left\{ \mathop{\rm Tr}_{\{\tau _\ell \}} \left[ \prod _{\ell \ne 1} 
B(\tau_\ell x_\ell)\right] \frac{\sqrt{1-q}}{X} 
Y_1 \right\} \times 
\nonumber \\
&& \left\{ \mathop{\rm Tr}_{\{\tau _\ell \}} \left[ \prod _{\ell \ne 2} 
B(\tau_\ell x_\ell)\right] \frac{\sqrt{1-q}}{X} 
Y_2 \right\} \; \; \;, 
\end{eqnarray}
in which we have posed $X=X(\mu,q)\equiv 1+\mu (1-q)$ and where
\begin{equation}
Y_i \equiv 
\exp \left(-\frac{x_i^2\mu q}{2 X}\right) \left[ \frac{\tau_i x_i \sqrt{q}}{\sqrt{X}\sqrt{1-q}}
H\left(\frac{-\tau_i x_i \sqrt{q}}{\sqrt{X }\sqrt{1-q}}\right) +
\frac{1}{\sqrt{2\pi}}\exp \left(-\frac{x_i^2 \mu q}{2 X
(1-q) }\right) \right] \; \; \;.
\end{equation}
In the large $K$ limit, the asymptotic expressions of the order parameters
are $\mu \simeq \alpha^2 K^2$ and $q\simeq 1 - 128/\pi^2/\alpha^2$. 
Using the previous expressions of $\hat \Lambda_{011}, \Lambda_{011},
\overline{\Lambda}_{011}$, we have
\begin{equation}
\Delta _{011} (\alpha , K) \simeq \simeq {\sqrt{2} \over \pi ^3 K}
\end{equation}
when $K\gg1$ and $\alpha\gg1$. Therefore, our RS solution is unstable against 
011 replicon fluctuations. However, in the large $K$ limit,
$\Delta_{011}$ vanishes and the RS Ansatz becomes marginally stable.

Let us now analyse the (122) mode. Similar calculations lead to
\begin{eqnarray}
\hat \Lambda _{122} &=& \frac{1}{\mu ^2 K} r^2  \nonumber \\
\Lambda _{122} &=& \frac{\alpha \mu ^2}{r^2} \int \prod_{\ell=1}^K Dx_\ell
\frac{N_5}{D} \nonumber \\
\overline{\Lambda}_{122} &=& 0 
\label{122}
\end{eqnarray}
where 
\begin{equation}
N_5 = \mathop{\rm Tr}_{\{\tau _\ell \}} \left[ \prod _{\ell=2}^K 
B(\tau_\ell x_\ell)\right] \frac{\exp \left(-x_1^2\frac{\mu q}
{2 (1+\mu(1-q))}\right) }{\sqrt{1+\mu (1-q)}}
H\left(\frac{-\tau_1 x_1 \sqrt{q}}{\sqrt{1+\mu (1-q)}\sqrt{1-q}}\right)
\; \;. 
\end{equation}

Therefore, we obtain
\begin{equation}
\Delta _{122} ^{(Com)} (\alpha , K) \simeq - {1 \over 2 K^2}
\end{equation}
when $K\gg1$ and $\alpha=O(1)$. We notice that the 122 mode is always
stable and a unique order parameter $q_*$ is thus sufficient to
describe the volume associated to a set of internal representations
${\cal T}$. 

\section{Analysis of ${\cal M}_D$ in the $K>>1$ Limit}

Let us now turn to the generalization problem. The Bayesian entropy
$-s(r=1)$ is given by
\begin{equation}
S_G =\mathop{\rm Extr}_{q} \left\{ \frac q2 + \frac{1}{2} \ln (1-q)+2\alpha 
\int \prod _\ell Dx_\ell {\cal H}(\{x_\ell \}) \ln {\cal H}(\{x_\ell \})
\right\}
\end{equation}
where, as $r=1$, $ {\cal H}(\{x_\ell \})$ depends on $q$ only (\ref{Tr}).
In the large $K$ limit, the scaling of $q$ is
\begin{equation}
q \simeq 1- \frac{\pi^6 \Gamma^4}{2\alpha ^4}
\end{equation}
where $\Gamma$ has been defined in (\ref{Gam}). 
Therefore, the Bayesian entropy asymptotically equals $S_G=2\ln \alpha$
and the generalization error decreases as $e_g = 2 \Gamma /\alpha$.
This proves that, contrary to the parity machine case\cite{watkin,opper},
only a small fraction among the $2^P$ possible sets of outputs contribute
to $S_G$ and explains why the generalization curve is smooth around 
$\alpha_D \sim \sqrt{\ln K}$ \cite{hertz}, defined by an average over all
sets of outputs. 
The typical entropy of internal representations is given by
\begin{eqnarray}
{\cal M}_D &=& \mathop{\rm Extr}_{q^*} \left\{ - \frac 12 \ln(1-q^*)
+ \frac 12 \ln (1-q) +\frac{1}{2} (q-q^*) + 
2\alpha \int \prod _\ell {\cal H}(\{x_\ell \}) \ln {\cal H}(\{x_\ell \})
-\right. \nonumber \\
&& \left. 2 \alpha K \int Dx H\left( x \sqrt{\frac{q^*}{1-q^*}} \right)
\ln H\left( x \sqrt{\frac{q^*}{1-q^*}}\right)\right\} 
\end{eqnarray}
In the limit $1\ll\alpha\ll K$, the internal overlap $q^*$ scales
as 
\begin{equation}
q^* \simeq 1- \frac{\pi^2 \Gamma^2}{2\alpha ^2 K^2}
\end{equation}
and the entropy of contributing internal representations reads
\begin{equation}
{\cal M}_D = \ln K - \ln \alpha 
\end{equation}
Therefore, $\alpha_D$ defined for the storage problem, and more generally
the Vapnik--Chervonenkis dimension \cite{vc}, are not relevant for
the typical generalization properties of a large committee machine
inferring a learnable rule. We can moreover note that above $\alpha_G
\simeq K$, one single domain survives and the generalization error
asymptotically decreases as $e_g = \Gamma /\alpha$ as is for finite $K$ and
large $\alpha$\cite{hertz,nosotros}. To end with, the condition $q=q^*$
signaling that a unique volume is non empty gives back the estimated
value of $\alpha_G$.

\section{Conclusion}

In this paper we have developed a complete analysis of the learning
and generalization properties of large committee machines.  Our
approach -- in which the weight space is partitioned according to the
internal representations of the learning examples -- allows us to
derive the relevant entropies ${\cal N}_R$, ${\cal N}_D$ and ${\cal
M}_D$ and successively to find the storage capacity of the model, to
verify the stability of the solution and to study the rule inference
capability.  In ref. \cite{nosotros} we have discussed the physical
and geometrical issues arising in the application of such a method to
the learning and generalization theory of MLN. Here the chief results
are the explicit derivation of the asymptotic storage capacity endowed
with a detailed analysis of the stability of the solution and the
derivation of the generalization cross--over at $\alpha=K$. From a 
methodological point of view, it is interesting to note that the RS
computation of the distribution of volumes is very close to the one-step
calculation of the Gardner volume. However, it is technically simpler and
allows for instance the derivation of the asymptotic storage capacity
of large committee machines while the same quantity seemed out of reach 
using the standard RSB computation\cite{bhs}.


\vspace{.3 truecm}

\begin{references}
\bibitem[*]{rm}  Email: monasson@physique.ens.fr

\bibitem[\dagger]{rz}  Email: zecchina@to.infn.it

\bibitem{nosotros}  R. Monasson and R. Zecchina, {\em Phys. Rev. Lett.} {\bf %
75}, 2432 (1995)

\bibitem{rete}  R. Monasson, D. O'Kane, {\em Europhys. Lett.} {\bf 27}, 85
(1994)

\bibitem{hertz}  H. Schwarze, J. Hertz, {\em Europhys. Lett.} {\bf 20}, 375
(1992)

\bibitem{gard}  E. Gardner, {\em J. Phys.} {\bf A} 21, 257 (1988) \hfill %
\break \hskip .4cm E. Gardner, B. Derrida, {\em J. Phys.} {\bf A} 21, 271
(1988)

\bibitem{bhs}  E. Barkai, D. Hansel, I. Kanter, {\em Phys. Rev. Lett.} {\bf %
65}, 2312 (1990) \hfill \break \hskip .4cm E. Barkai, D. Hansel, H.
Sompolinsky, {\em Phys. Rev. A} {\bf 45}, 4146 (1992)

\bibitem{mpv}  M. M\'ezard, G. Parisi, M.A. Virasoro {\em Spin Glass Theory
and Beyond} (World Scientific, Singapore, 1987)

\bibitem{kondor}
T. Temesvari, C. De Dominicis, I. Kondor, {\em J. Phys.} {\bf A} 27, 
7569 (1994)
\bibitem{corr}
Formulae (5-6) in reference \cite{nosotros} are not correct~: the power $r$
must appear in the argument of the logarithm only. However, all asymptotic
results derived in \cite{nosotros} are true when $K\gg 1$.
\bibitem{vc}
V.N. Vapnik, {\em Estimation of Dependences Based on Empirical Data}
(Springer-Verlag, New York, 1982)
\bibitem{opper}
M. Opper, {\em Phys. Rev. Lett.} {\bf 72}, 2113 (1994)
\bibitem{mit}
G.J. Mitchison, R.M. Durbin, {\em Bio. Cybern.} {\bf 60}, 345 (1989)
\bibitem{watkin}
T. Watkin, A. Rau, M. Biehl, {\em Rev. Mod. Phys.} {\bf 65}, 499 (1993)

\end{references}
\end{document}